# NOVEL FRAMEWORK FOR MOBILE COLLABORATIVE LEARNING (MCL) TO SUBSTANTIATE PEDOGOGICAL ACTIVITIES

[1]Abdul Razaque  [2] Khalid Elleithy  [3]N.salama
[1]arazaque@bridgeport.edu  [2]elleithy@bridgeport.edu  [3]nsalama@bridgeport.edu

Wireless and Mobile communications Laboratory
Department of Computer Science and Engineering
University of Bridgeport, USA

**ABSTRACT**

Latest study shows that MCL is highly focusing paradigm for research particularity in distance and online education. MCL provides some features and functionalities for all participants to obtain the knowledge. Deployment of new emerging technologies and fast growing trends toward MCL boom attract people to develop learning management system, virtual learning environment and conference system with support of MCL. All these environments lack the most promising supportive framework. In addition some of major challenges in open, large scale, dynamic and heterogeneous environments are not still handled in developing MCL for education and other organizations. These issues includes such as knowledge sharing, faster delivery of contents, request for modified contents, complete access to enterprise data warehouse, delivery of large rich multimedia contents (video-on-demand), asynchronous collaboration, synchronous collaboration, support for multi model, provision for archive updating, user friendly interface, middleware support and virtual support.

To overcome these issues; the paper introduces novel framework for MCL consisting of four layers with many promising functional components, which provide access to users for obtaining required contents from enterprise data warehouse (EDW). Novel framework provides information regarding the course materials, easy access to check the grades and use of labs. The applications running on this framework give substantial feedback for collaboration such as exchange for delivery of communication contents including platform for group discussion, short message service (SMS), emails, audio, video and video-on-demand to obtain on-line information to students and other persons who will be part of collaboration.

## 1. INTRODUCTION

Computer-supported collaborative learning (CSCL) is one of the most promising pedagogical paradigms, which supports science of learning and research trend in education. This paradigm is highly acknowledged and implemented in schools, colleges, universities and different organizations across the world. From other side; CSCL applications are very expensive and not affordable to deploy for many educational institutions and small types of organizations. Even most of the active communities are still completely unable to use this technology for delivering and stimulating the knowledge. In addition, busy life and hard schedule do not allow the people to use static technology. They want to obtain the learning materials in dynamic environment anytime and anywhere. Mobile is only attractive device with rich features that provides MCL in free environment to meet pedagogical objectives [1].

The concept of mobile-based learning is completely different from classroom-based learning method. Collaboration is synchronous and coordinated activity that creates the bridge of perpetual efforts to build and maintain a shared notion of a problem. MCL in education involves with joint intellectual efforts initiated by students or teachers. It can play phenomenal and increasingly important part in education. It creates an interaction and promotes awareness [4, 13]. MCL enhances critical analysis and helps the students to clarify their concepts from other students [12]. MCL is key factor of fostering brainstorming in groups because students with different thoughts on variety of possible variable can perform their activities in peers. It provides sufficient time to students for interaction in order to discuss and exchange the knowledge through higher degree of thinking. The major focus of MCL is to create cooperation instead of competition and take a task as challenge to accomplish as graphic organizer. Students also explore various new things and they are less reliant on teacher's feedback [6, 11].

MCL also exhibits intellectual synergy of various combined minds coming to handle the problems and stimulate the social activity of mutual understanding [2]. MCL contributes a larger pedagogical agenda. This approach provides many possibilities, such as increasing literacy rate, motivating the people to improve their education, capturing the market money and providing the opportunities to group of persons, working in same or different organizations [14]. Implementation of novel framework will enable educational institutions and other organizations to deploy at minimal cost. It replaces the class based learning and benefits the persons who are far from the educational institutions and other organizations. Students do not need to attend a class for just listening lectures and attending video conferences.

They can get the contents on mobile by just registering with server of educational institution or related organizations. The novel framework will motivate different walks of people to establish social networks, community group, law enforcement agencies, and small groups of security force to ensure the safety of the sensitive places of city. It will also combine various branches of defense department for coordinating with troops for special causes. It helps the military for launching new projects and sharing strategic based information about any secret plans. Novel framework can substantiate health department for introducing new health projects to share the information and create awareness.

## 2. RELATED WORK

Xiaoyong Su et al. [5] propose the four layer framework for multimedia content generation and prototype for multimedia mobile collaborative system. The proposed framework gives an idea how to handle user, device and session management. They also suggested that mobile collaborative environment could be possible by upgrading the devices and network technologies. Lahner F and Nosekabel H [8] have implemented the program in University of Regensburg, Germany, which supports e-learning contents to be displayed on mobiles. The structure of system provides the facility to users to get same contents via mobiles. Authors claim [15] that their proposed approach will improve academic and learning activities. The prototype is based on user profile which stores information regarding learning process. The second is location system, which is used to identify the physical location of user supported by generic architecture. Third is personnel Assistant (PA), which resides in the mobile. Fourth is learning object repository, which stores the contents, related with the process of pedagogical teaching and fifth is message sending system to be controlled automatically or using administrative interface by operators. Final component is Tutor that searches learning opportunities. The proposed prototype provides idea of online learning.

Stanford University [7] launched pilot project the Dunia Moja project"one world" in Swahili by July 5, 2007 with three partner universities in South Africa, Tanzania and Uganda. The purpose of this project was to introduce web-based courses and materials. Sony Ericson and Ericsson provided free mobile smart phones. The focus of the project was to use the mobiles to access the course-web site and send text messages.

Allision Druin et al. [10] have discussed the prototype for their ongoing participatory design project with intergenerational design group to create mobile application and integrate into iP Phone and iPod touch platforms. Authors claim that designed application can provide the opportunities to bring the children and grand parents together by reading and editing the books. Ericsson Education Dublin [9] initiated a project with €4 million and focused on e-learning to m-learning. The main objectives of project were to produce a series of courses for mobiles, PDAs and smart phones. All of these studies in MCL environment show that neither any contribution particularly worked on architecture based design and nor on functional systems to support mobile application efficiently. All of the previous survey mostly developed the courses for mobiles and few contributions touched the technology but not implemented. Novel framework supports to server and client side systems with promising features to meet the pedagogical requirements. Novel framework is handy in introducing rich MCL client-server based management systems, virtual learning environments and conference systems within educational institutions and outside of the institutions.

The challenging issues motivating this research are: to what degree novel framework is effective and result oriented? Can it handle all important features necessary for effective MCL? Can this novel framework support to mobile collaborative applications and other software threats sufficient to meet pedagogical needs and other collaborative requirements? How the findings can be used to engineer the requirements of Students, teachers and other administrative staff of UB and other institutions? What are socio-economic impacts of the novel framework on USA and beyond the world whether claimed objectives have been achieved or not if yes then to what extend and how? The goal of this study is to obtain the learning materials in order to meet pedagogical needs and other collaborative requirements of educational institutions including different organizations.

## 3. NOVEL FRAMEWORK FOR MCL

To make successful collaboration; we need organized architecture with support of latest technologies to meet our expectations. Various conceptual collaborative architectures have been proposed so for. Xiaoyong Su et al. [5] proposed four layered components for collaborative framework, which consists of content generation layer, communication layer, content regeneration layer and content visualization layer. Each layer has been assigned different responsibility. The architecture of [5] has been optimized and extended with inclusion of new sub components. Figure:1 explains the novel framework for mobile collaborative learning environment.

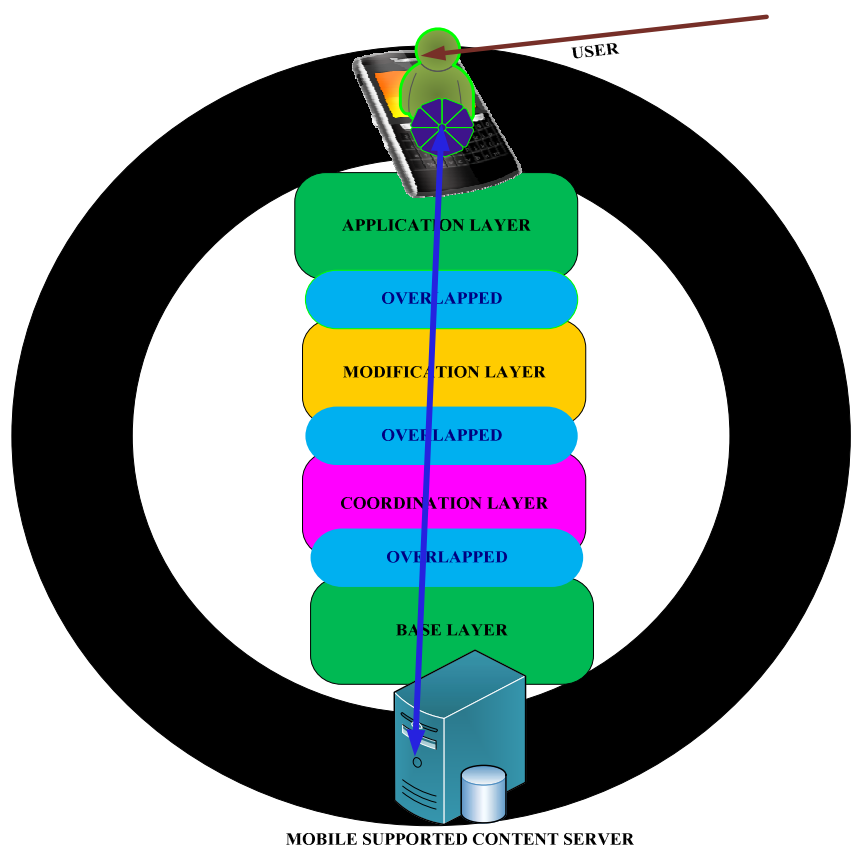

Figure 1: Conceptual Architecture to generate the content for MCL

### 3.1 BASE LAYER

The content generation layer is main component of collaborative framework. The Figure 2 shows the working process of this layer. If client requires any contents, sends the request message to content server for delivery of required contents. The request message includes device profile, status of previous network condition and required URL. The mobile information device profile (MIDF) is supported with Java platform. J2ME is standard model for mobile technology. The base of J2ME is on three different layers, which are profiles, Java virtual machines and configurations. MIDP is based on Java runtime environment supported with profiles and configuration layers. These both layers sit on operating system of client's mobile and personnel digital assistants (PDAs). The goal of profile is to represent the application programming interface (API) for same type of mobile devices with similar features. From other side, content generation layer has also support of different programming language such as C++, Java, HTML, XML, Flash, XHTML, and Structured Query Language (SQL). When content server gets request from client, it searches the requested contents into enterprise data warehouse; if requested content is found then forwarded to message server for further process.

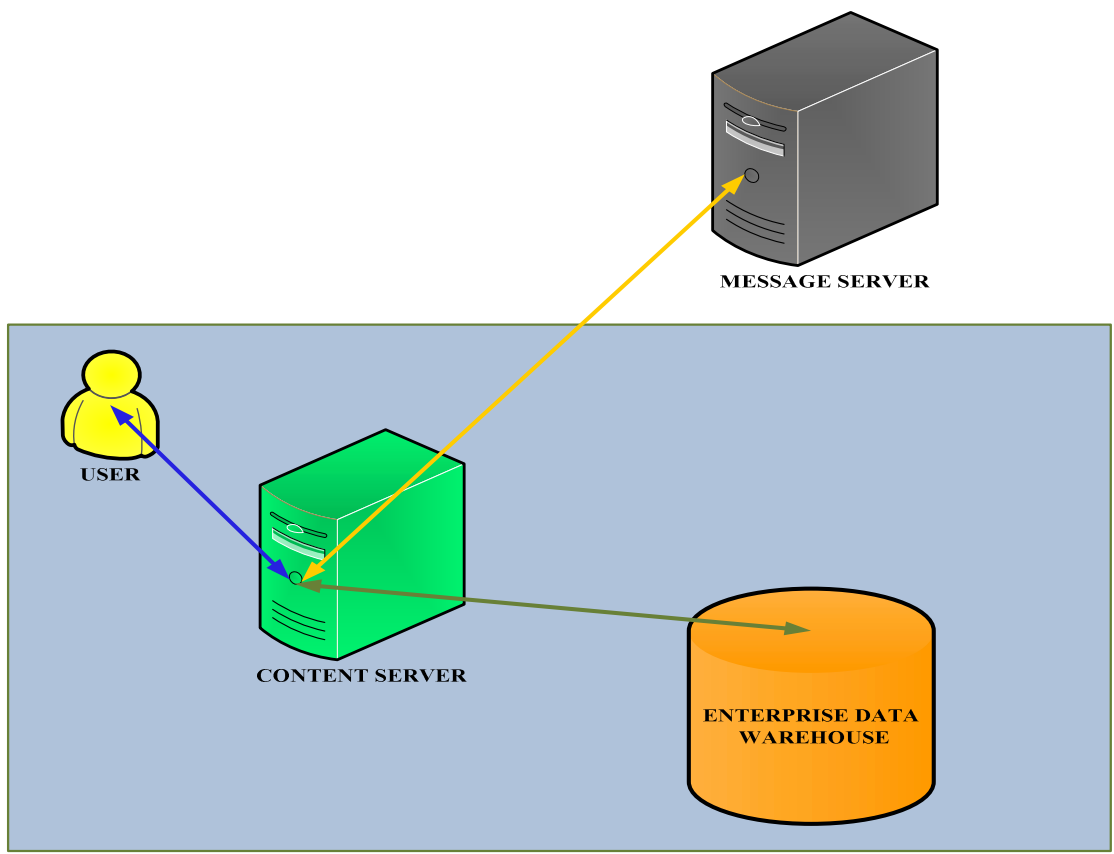

Figure 2: Process of base layer in mobile collaborative environment

## 3.2 COORDINATION LAYER

This layer works as transport layer. When message server gets requested content from base layer; first checks network connection. If network connection is enabled, then checks the size of requested content. If requested content contains small amount of data, forwards the content to content integrating manager (CIM) to next layer. If size of requested content is large, then sends the content to message forwarding manager (MFM) for fragmentation of the message. MFM fragments the message into chunks (small amount of data) and transports to (CIM). MFM continuously keeps on transporting the chunks until last chunk means whole fragmented message is delivered to (CIM). In case, if network is disabled then message server (MS) has also option to send the messages (content) to message manager (MM). When MM gets message from MS, it stores the messages into the buffer message. When network connection retains, MM forwards the stored messages to MS and figure 3 shows working process of coordination layer.

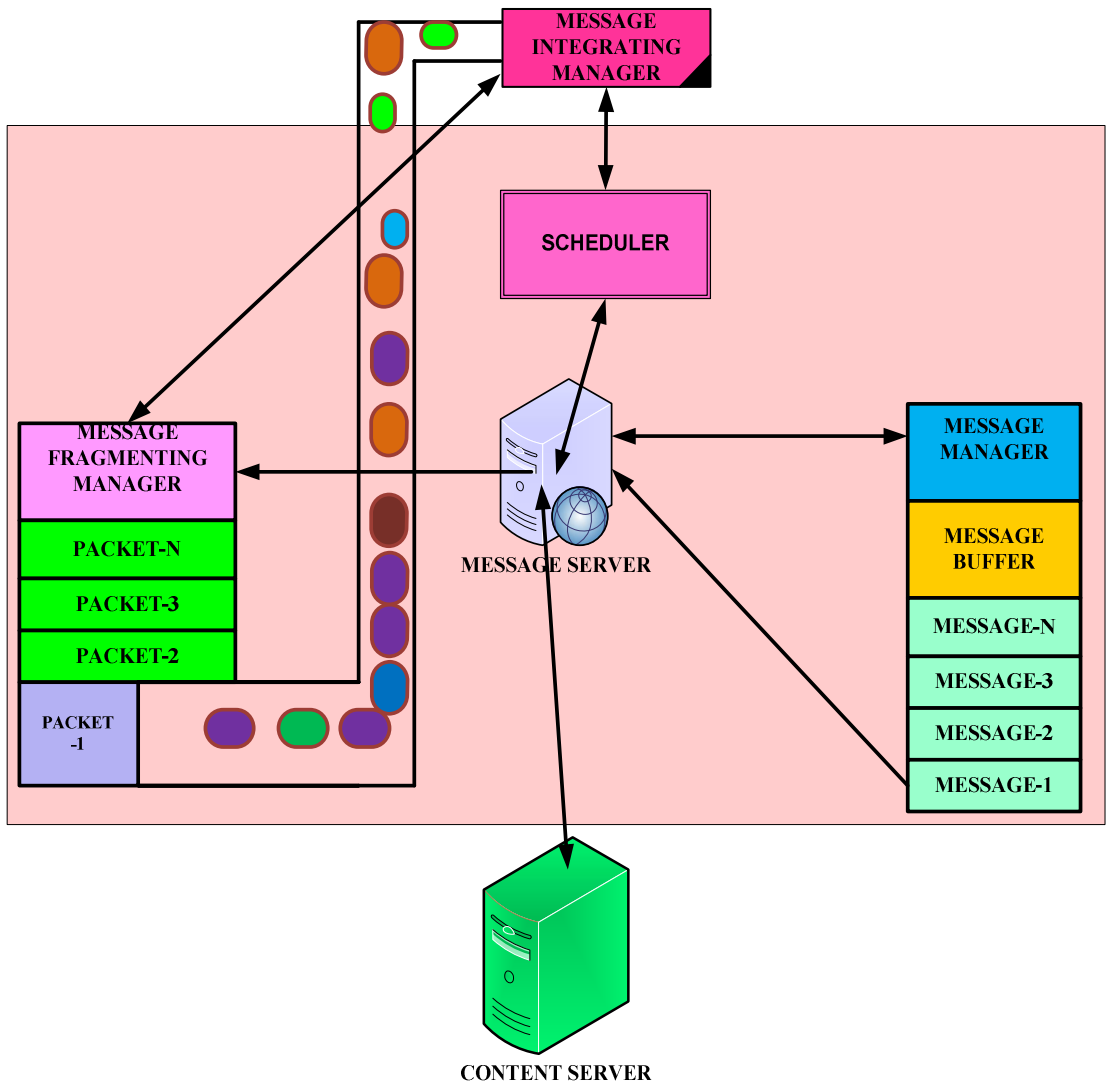

Figure3: Working Model of coordination layer in mobile collaborative environment

## 3.3 MODIFICATION LAYER

This layer performs two types of tasks; first, it forwards the content to application layer for next process. Second, if client sends modification request for content in middle of the process, then also starts to work on modified request. The process of this layer starts from CIM. The CIM has dual functionalities; it integrates fragmented chunks received from MFM of previous layer. The CIM forwards integrated chunks in form of messages (content) to CFM. CIM also receives content from MIS that comprises of small amount of data. When CFM gets content from CIM, it forwards to media manager (MM), in case no request for content modification is received from Client. If content modification request is received, CFM sends the content to content modifier manager (CMF) with requested modification. The CMF sends content to content deciding manager (CDM) in both conditions.

If CMF modifies the content, then sends modified content to content deciding manager (CDM). If modified content meets request of client, CDM forwards requested modified content to CFM for further process. Otherwise it sends back to CFM to same layer. CDM also deals with same process with original content and gets back to CFM. The CFM fragments the contents in chunks and forwards to CIM of previous layer shown in figure 4.

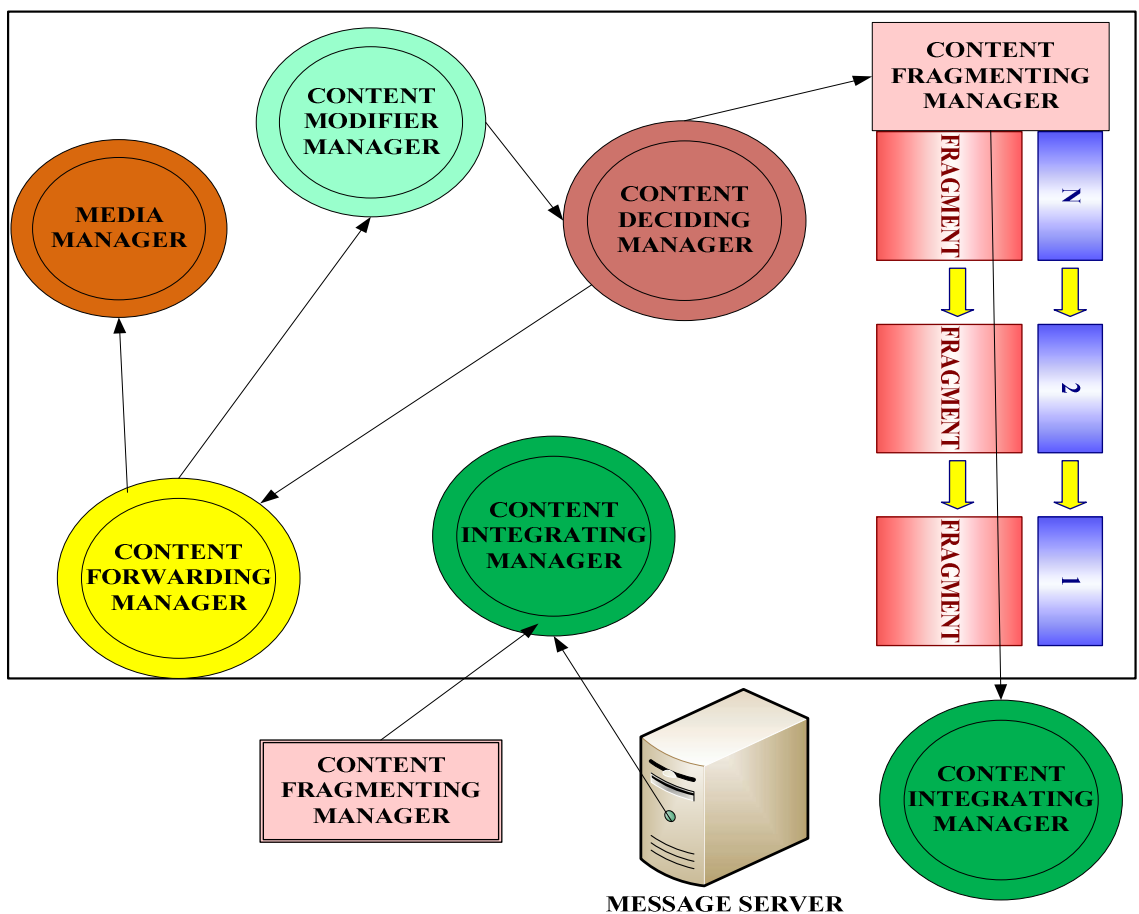

Figure 4: Modification process in mobile collaborative environment

## 3.4 APPLICATION LAYER

The function of this layer is to display and hand over the contents to client. Application layer obtains contents from media manager and displays the contents in form of data, graph, image and voice as per request of client. Actually media manager is a tool, which organizes files to be shared between mobile devices. The main function of layer is to translate the source program into object program that is done with support of parse engine. The used engine is faster and robust to handle the data errors. Parsing includes set of components, which are scanner component, parser component and document type definition (DTD) component.

Scanner component is first major component of parsing engine, which provides push-based API. An API gets characters from input stream (mostly URL) and obtains proper sequences then remove useless data by using methods. Parser component coordinates and handles activities of other components of system. DDT explains the tags, associated set of attributes and hierarchical rules for well defined and valid documents in form of target grammar. Parse engine requires five phases to transform multimedia contents to document object model (DOM). Before delivering document to requested client, each document passes from object construction phase, opening input stream phase, tokenization phase, token iteration phase and object destruction phase. In object construction phase, an appropriate application starts process by creating URL, tokenizer objects and language tags. Since the parser is provided sink and DTD. The function of DTD is to understand the grammar of documents. Sink provides interface for DTD to make content and model properly.

Opening an input scream phase gets URL in shape of network input stream. The parse engine provides input stream to Scanner, which controls whole process. This process continues till End-OF-File (EOF) occurs. Token iteration phase validates documents and builds content model. Finally object destruction phase destroys the objects that are in parse engine after tokenization and iteration process. Application layer parses document with support of parse engine to display an object. Simple process of parse engine is shown in figure 5. The following limitations of parse engine has some resolved.

Netlib (Input Scream and URL); The XP_COM systems: nsString; nsCore.h and prototypes.h

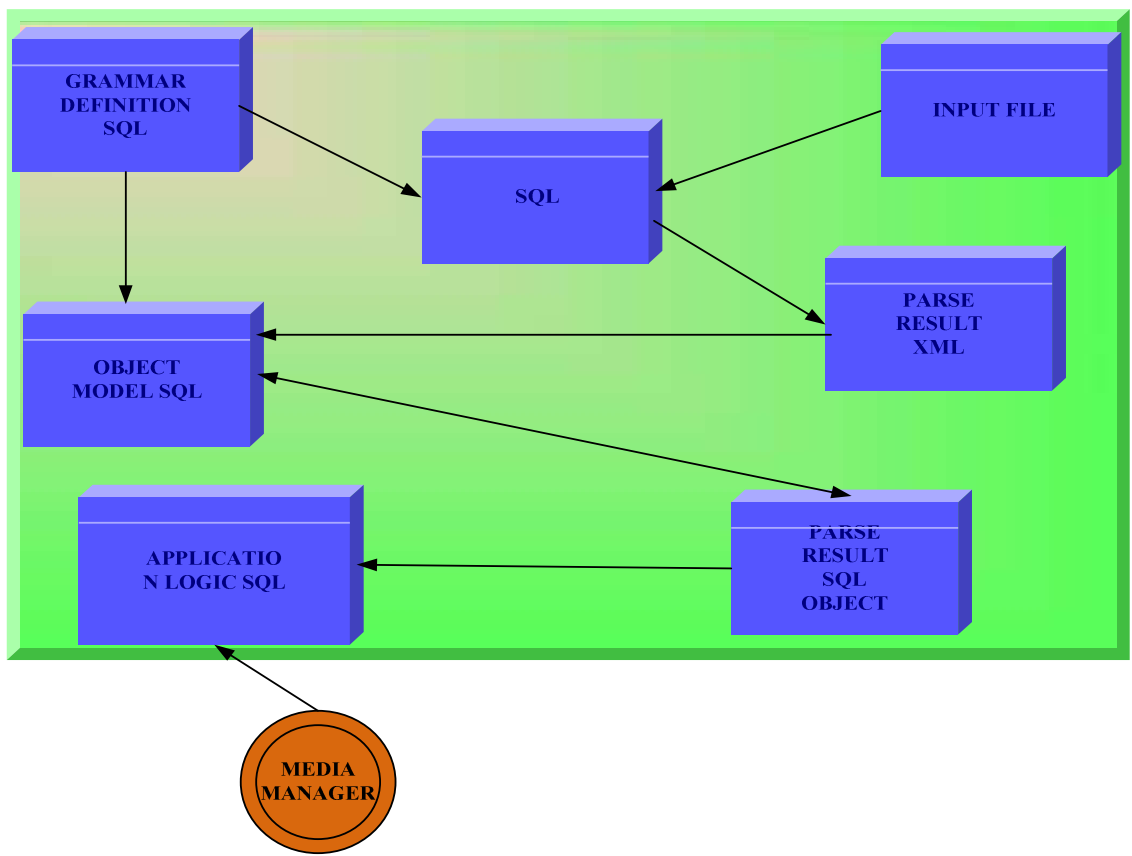

Figure 5: Simple working model of optimized parse engine in mobile collaborative environment

To validate the novel framework; application is run on it that is developed with combination of SDK and android for delivery of text, images, graphs and voice media contents in form of the message. The function of message service is to mange and delivers the message. The message server has two parts; message agent and message server. Profile management is sub component, which falls within content generation, delivery and coordination components. This handles profile of mobile users. A mobile user provides the information regarding the hardware, resource, operating system and connection status. The system architecture of collaboration consists of four logical parts in [3]. The four parts includes with infrastructure, service function, advance service function and application. These logical parts provide different functionalities; infrastructure includes operating systems, communication modules; service function consisting of floor management, media management and session management. Advance service function provides creation and deletion of shared and video window.

## 4. CONCLUSION

We conclude that novel framework has made significant progress in design and development of the MCL environment. First, the paper presents novel framework with integration of new components, such as Content integrating manager, content deciding manager, content modifying manager and content fragmenting manager. These components collectively protect loss of data and make the fast and robust process of delivery of contents to users. Second, we have implemented the first layer of the novel framework to obtain the learning materials on hand held devices particularly on mobile devices. With implementation of framework, students get the contents of the courses anywhere and anytime. We have made significant modifications in the four layer architecture previously explained in [5]. Our framework provides an efficient and faster way of delivering the contents to mobile node. If any modified request is made by client side that is also smoothly handled on third layer rather than going to first layer. These characteristics of novel framework give fast provision of content to mobile node. In future, we will test our new group application with rich features on this novel framework.

# REFERENCES

[1] Abdul Razaque and Khaled Elleithy "Architecture based Prototypes for Mobile Collaborative Learning to improve Pedagogical activities" Interactive collaborative learning (ICL), IEEE, Piešany, Slovakia, September 21 – 23, 2011.

[2] Abdul Razaque and Khaled Elleithy, "Interactive linguistic prototypes to foster pedagogical activities through mobile collaborative learning (MCL)"International Journal of Interactive Mobile Technologies (iJIM)",Vol 5, No 3 (2011).

[3] Jooung-Souck Sung,"Application Protocol Design for Collaborative Learning", International Journal of u-and e-service,science and technology Vol.2, No.4,December, 2009.

[4] .Levent Gorgu, Gregory M.P. O'Hare, Michael J.O'Grady,"Towards Mobile Collaborative Exergaming",IEEE second international conference on Advances in Human-oriented and personalized mechanisms, Technologies, and services,2009,pp 61-64.

[5] Xiaoyong Su, B.S.Prabhu, Chi-Cheng Chu, Rajit Gadh, "midleware for multimedia mobile collaborative system", IEEE third annual wireless telecommunications symposium (WTS, 2004), May 14-15, 2004, Calpoly pomona, Clofornia, USA.

[6] Jacobs, G. M., & Hannah, D. (2004). Combining cooperative learning with reading aloud by teachers. International Journal of English Studies, 4, 97-118.

[7] http://www.the-icsee.org.

[8] Lahner F,Nosekabel H, " The Role of Mobile Devices in E-Learning First Experiences with a wirelss E-Learning Envrronment, Proceedings of the IEEE International Workshop on wirelss and mobile technologies in Educatio, August 29-30, 2002, Vaxjo, Sweeden.

[9] http://learning.ericsson.net/mlearning.

[10] Allison Druin, Benjamin B.Benderson, Alex Quinn,"Designing Inter generational Mobile Storytelling" IDC 2009-Workshops, 3-5 June, 2009, Como, Italy.

[11] Sergio Martin, Ivica Boticki, George Jacobs, Manuel Castro and Juan Peire," Work in Progress - Support for Mobile Collaborative Learning Applications, 40[th] ASEE/IEEE Frontiers in Education Conference, 2010, Washington, DC.

[12] The Metiri Group,"The Impact of Collaborative, Scaffolded Learning in K-12 Schools: A Meta-Analysis", Cisco Systems,2009.

[13] .Gene Chipman, Allison Druin, Dianner Beer, Jerry Alan Fails, Mona Leigh Guha and Sanate Simms, "A case study of Tangible Flags: A collaborative technology to enhance field trips", University of Maryland, Human computer International Lab college, MD, USA.

[14] Kwang Lee and Abdul Razaque,"Suggested Collaborative Learning Conceptual Architecture and Applications for Mobile Devices", springer,Lecture Notes in Computer Science, 2011, Volume 6769/2011, 611-620.

[15] Jorge Barbosa, Rodrigo Hahn, Debora N.F.Barbosa, Claudio F.R.Geyer," Mobile and Ubiquitous Computing in an Innovative Undergraduate Course", SIGCSE07, March 7-10, 2007, Covington, Kentucky, USA.